\documentclass[12pt,preprint]{aastex}

\slugcomment{Accepted for publication in ApJ (Part 1)}

\shorttitle{Polarization of L Dwarfs}
\shortauthors{Sengupta and Kwok}
\begin{document}

\title{Polarization of L Dwarfs by Dust Scattering}  

\author{Sujan Sengupta}
\affil{Institute of Astronomy and Astrophysics, Academia Sinica, P.O.Box 23-141,
       Taipei 106, Taiwan\\
 and \\ Indian Institute of Astrophysics, Koramangala, Bangalore 560 034, India}
\email{sujan@iiap.res.in}
\author{Sun Kwok}
\affil{Institute of Astronomy and Astrophysics, Academia Sinica, P.O.Box 23-141,
       Taipei 106, Taiwan}
\email{kwok@asiaa.sinica.edu.tw}

\begin{abstract}
The degree of polarization in L dwarfs of spectral types L0 to L6 resulting
from dust scattering in a rotation-induced oblate photosphere is calculated.
Assuming that forsterite is the main condensate, the atmospheric dust
distribution is derived for different spectral types based on a chemical 
equilibrium model.  The degree of polarization at optical is then calculated
using a single scattering model.  The expected linear polarization at optical
is found to peak at around spectral type L1. For a fixed rotational
velocity, the degree of polarization decreases from hotter to cooler objects.
However, with the increase in mean grain size, the degree of linear
polarization reduces significantly. We fit the recently observed linear
polarimetric data of L dwarfs and find that single dust scattering model
coupled with the chemical equilibrium models of condensates is consistent
with the observational results.

\end{abstract}


\section{Introduction}

Observations of L dwarfs with effective temperatures of 1400-2200 K have
led to the investigations of dust condensates in their atmospheres
\citep{kir99}.  Because of complete gravitational settling, grains are
expected to condense beyond the visible atmosphere for objects with effective
temperatures below 1400 K.  At higher effective temperatures, grains can be
present in the visible atmosphere due to incomplete gravitational settling 
\citep{bur99, bur01, ack01, all01, tsuji04, tsuji96}.

Using a thermodynamical model based on homogeneous and heterogeneous 
condensation assumptions, \citet[hereafter C03]{coo03} were able to
determine the distribution of different species of dust particles.
Using solutions of dust moment equations in a static atmosphere, 
\citet{woi03, woi04} studied the continuous nucleation of solid particles
from the gas phase.  By imposing the requirement of minimum mixing activities
and thermodynamical stability, they were able to determine the upper boundary
(called ``cloud deck'') and the lower boundary (called ``cloud base'')
respectively.

The possibility of detecting polarization at optical from grains in the 
atmospheres of
L dwarfs was first raised by \citet{sen01}.  Since fast rotation of L dwarfs
\citep{bas00} will induce the shape of their photosphere into the form of an
oblate ellipsoid, this non-sphericity will lead to incomplete cancellation of
the polarization from different areas of the stellar surface.
This prediction was first confirmed by the detection of linear polarization at
768 nm from a few L dwarfs by \citet{men02}. Recently \citet{oso04}, 
have reported R and I band detection of linear
polarization from several L dwarfs. Since polarization in the optical
is unlikely to be due to Zeeman splitting of atomic or molecular lines or
by synchrotron radiation, the observed polarization 
can best be explained by single dust scattering in a rotationally induced
oblate atmosphere \citep{sen03}.

The observed polarization of DENIS-P J0255-4700 and 2MASSW J0036+1821 was
modeled by \citet{sen03} by assuming minimum oblateness of these objects
based on a model of complete degeneracy of a non-relativistic polytropic gas
under hydrostatic equilibrium.  The observed degree of polarization was 
obtained by considering different parameters for the scale height, number
density, and mean radius of the grains.  By minimizing
the oblateness and the number density of grains, an upper limit on the mean
grain radius that is consistent with the polarization observed by \citet{men02}
was obtained.
However, the process of dust condensation was not considered.  

In the present paper, we present detailed theoretical models of optical linear
polarization from L dwarfs of spectral types ranging from L0 to L6 based
on the dust condensation  model of C03.  With this model, we fit the observed
polarization from several L dwarfs and predict the amount of 
polarization expected from any L dwarfs with given rotational velocity.

In section 2, we present the formalism for the calculation of polarization
from single dust scattering in an oblate medium.  In section 3, the adopted
atmospheric model is described.  The estimation of rotationally induced
oblateness is presented in section 4, and the dust model parameters are
described in section 5.  The results and discussions are presented in
section 6, followed by conclusions in section 7.

\section{Polarization by single dust scattering} 

If a stellar object is perfectly spherical then the net polarization would
be zero due to the cancellation of the contribution of each point on the
photosphere. The observation of non-zero polarization from a few L dwarfs
therefore suggests that the scattering geometry is asymmetric, which could be
the result of fast rotation of the objects. 
Since the dust density is low and scattering by atoms and molecules does
not contribute to polarization significantly, single scattering approximation
is reasonable for the region where the optical depth $\tau < 1$. 
If present, multiple scattering can reduce the degree of polarization by a 
few orders of magnitude \citep{sen01} because the planes
of the scattering events are randomly oriented and average each other's
contribution out from the final polarization.  The effects of oblateness of 
an object on linear polarization due to single scattering by spherical grains
have been discussed by \citet{dol95}  and by \citet{sim82}.  In this paper,
we adopt the formalism given by \citet{sim82}. 

For an optically thin atmosphere, the total linear polarization $p(k)$ can be
written as
\begin{equation}
p(k)=|Z(k)|=|Z^*(k)|,
\end{equation}
where $k=2\pi/\lambda$, $\lambda$ being the wavelength, the asterisk denotes
the complex conjugate and
\begin{eqnarray}\label{abc}
Z(k)=\frac{1}{k^2}\int\int\int\frac{i_1(\theta,k)-i_2(\theta,k)}{2}n(r,\theta,\phi)
e^{2i\phi} d\omega dr.
\end{eqnarray}
where $dw$ is the element of solid angle.
In the above equation, $\theta$ is the scattering angle, $n$ is the number
density of scattering particles, $i_1$ and $i_2$ are the
scattering functions given by \citep{van57}
\begin{eqnarray}
i_1(\theta)=\left|\sum^{\infty}_{n=1}\frac{2n+1}{n(n+1)}[a_n\pi_n(\theta)+b_n\tau_n(\theta)]
\right|^{2},
\end{eqnarray}
and 
\begin{eqnarray}
i_2(\theta)=\left|\sum^{\infty}_{n=1}\frac{2n+1}{n(n+1)}[b_n\pi_n(\theta)+a_n\tau_n(\theta)]
\right|^{2},
\end{eqnarray}
where
\begin{equation}
\pi_n(\theta)=-\frac{1}{\sin\theta}P^1_n(\cos\theta),
\end{equation}
and
\begin{equation}
\tau_n(\theta)=-\frac{d}{d\theta}P^1_n(\cos\theta).
\end{equation}
The co-efficients $a_n$ and $b_n$ are in general complex functions of the
refractive index $m$ and the particle radius to wavelength ratio.

For a smooth density distribution, one can write
\begin{equation}\label{abc1}
n(r,\theta,\phi)=\sum^{\infty}_{l=0}\sum^{m=l}_{m=-l}n_{lm}(r)Y_{lm}(\theta,\phi),
\end{equation}
where 
\begin{equation}
Y_{lm}(\theta,\phi)=\alpha(l,m)P^m_l(\cos\theta)\exp(im\phi)
\end{equation}
\begin{equation}
\alpha(l,m)=\left[\frac{(2l+1)(l-m)\!}{4\pi(l+m)\!}\right]^{1/2},
\end{equation}
and $P^m_l$ is the associated Legendre function of the first kind. 

Substituting eq.~\ref{abc1} into eq.~\ref{abc} and integrating over
$\phi$, we get
\begin{equation}\label{abc2}
p(k)=\frac{2\pi}{k^2}\sum^{\infty}_{l=2}N_{l2}F_{l2},
\end{equation}
where
\begin{equation}
F_{lm}=\alpha(l,m)\int^1_{-1}\frac{i_1-i_2}{2}P^m_l(\cos\theta)d(\cos\theta).
\label{abc2a}
\end{equation}

Considering an axisymmetry density distribution with a rotational invariance
around some axis and using the addition theorem of spherical harmonic,
$N_{lm}$ can be written as
\begin{eqnarray}\label{abc3}
N_{lm}=2\pi\alpha(l,m)P^m_l(\cos \theta_i) e^{-2i\phi} \int^{R_1}_{R_2}n(r)dr\int^{1}_{-1}
\frac{P_l(\mu)d\mu}{[1+(A^2-1)\mu^2]^{1/2}},
\end{eqnarray}
where $R_1$ and $R_2$ are the outer and the inner equatorial axis length,
$A$ is the ratio of the length of the equatorial axis to the polar axis,
$\theta_i$ is the viewing angle of the observer, and $\mu=\cos\theta$.

At an edge-on view, $\theta_i=\pi/2$ and $\phi=0$, and hence $N_{lm}$ is real.
From the equation of  hydrostatic equilibrium, we have an expression of the 
particle density as a function of gas pressure:

\begin{equation}
n(r)dr=\frac{n(P)dP}{g\rho(P)},
\end{equation}
where $P$ is the pressure at different geometrical height, $\rho$ is the mass 
density at different pressure scale, and $g$ is the surface gravity (which
can be assumed to be constant for a geometrically thin atmosphere). 
Substituting eqs.~\ref{abc2a} and \ref{abc3} into eq.~\ref{abc2}, we have 

\begin{eqnarray}\label{pol}
p(\lambda)=\frac{\lambda^2}{g}\int^{P_2}_{P_1}\frac{n(P)dP}{\rho(P)}\sum^{\infty}_{l=2}
\left\{\alpha^2(l,m)P_l^m(0)F_{l2}\int^1_{-1}\frac{P_l(\mu)}{[1+(A^2-1)\mu^2]^{1/2}}d\mu
\right\}
\end{eqnarray}
for the degree of polarization for a hydrostatic atmosphere viewed edge on.

\section{The atmospheric models}
  
The effective temperature of the L dwarfs of different
spectral type can be approximated by the linear empirical relationship 
of \citet{ste01}:
\begin{eqnarray}
T_{eff}=2220-100\times L 
\end{eqnarray}
where $L$ is the spectral type between L0 and  L8.
Recently \citet{vrba04} and \citet{goli04} have presented 
$T_{eff}$ measurements for L and T dwarfs based on bolometric luminosities.
In the present work we adopt the sixth order polynomial fit given by
\citet{goli04} for translating the $T_{eff}$ into spectral
type. Fig. 1 shows the $T_{eff}$ for different spectral types calculated
by using the linear relationship given by \citet{ste01} and by the polynomial
formula provided by \citet{goli04}.
 The $T_{eff}$ calibration of \citet{vrba04} and \citet{goli04} agree well
in the interval L3-L8 but there
are significant differences in earlier types. 
In our calculations for the degree of polarization, the effective
temperature $T_{eff}$ is used and hence the degree of polarization should
be considered strictly as a function of $T_{eff}$ rather than the spectral
type.  

The surface gravity of L dwarfs older than about a few hundred million years
varies from $g=10^5$ cm s$^{-2}$ to $g=3\times10^5$ cm s$^{-2}$. Evolutionary
models by Chabrier et al. (2000) show younger L dwarfs to have surface
gravity smaller than $10^5$ cm s$^{-2}$. In this paper, we assume $g=10^5$
cm s$^{-2}$.

Formation of dust makes it a prohibitive task to develop a fully consistent
atmospheric model for ultra-cool dwarfs. This is mainly because of the fact
that the presence of dust cloud affects the radiative equilibrium of the
upper atmosphere and hence alters the T-P profile from that of a cloud-free
atmosphere. On the other hand, the T-P profile dictates the position and
the chemical equilibrium of condensates. Allard et al. (2001) presented
atmospheric models for two limiting cases, e. g., one with inefficient
gravitational settling wherein the dust is distributed according to chemical
equilibrium predictions (AMES-dusty) and another with efficient gravitational
settling in which situation dust has no effect on the thermal structure
(AMES--cond). \citet{tsuji04} have proposed a Unified Cloudy Model (UCM)
in which the segregation of dust from the gaseous mixture takes place in all
the ultra-cool dwarfs and at about the same critical temperature.

In the present work, the temperature-pressure profiles for  the L dwarfs with
different spectral type are calculated by solving the non-LTE radiative
transfer equations coupled with the hydrostatic equilibrium equations. 
We have employed the full Mie theory, incorporating
the dust opacity as well as the Mie phase function. The calculations are first
performed by taking only gas opacities.
We have adopted the atmospheric opacity sources discussed in \citet{sau00}
and assumed solar metallicity. The T-P profile thus obtained determined the
position of the cloud location. We then incorporated dust opacities to
calculate the final set of T-P profile. This T-P profile is then used again to
determine the base and the deck of the cloud.
Our pressure-temperature profiles are checked against those
presented in \citet{bur01} and in C03, as well as models kindly provided by
M. Marley (private communication). Nevertheless, the polarization profile
would change depending on the chemical equilibrium procedure adopted in
different atmospheric models.

\section{Rotation induced oblateness}
The observation of non-zero polarization from unresolved objects indicates the
non-sphericity of the photosphere. The non-sphericity may results from several
causes.  For example,  rotation of a stellar
object will result in a shape of an oblate ellipsoid, as is evident in the
outer solar planets.  At 1 bar pressure level, the eccentricity of Jupiter,
Saturn and Uranus are 0.35, 0.43 and 0.21 respectively.  Apart from rotational
effects, tidal interaction with the companion in a binary system also 
imposes an ellipsoidal shape extending toward the companion.

Spectroscopic studies by \citet{bas99} and \citet{bas00} indicate rapid
rotation of ultra-cool dwarfs along their axis.  The brown dwarf Kelu 1 is found
to be the fastest rotator with a  projected angular velocity ($v\sin i$)
as high as 60 km s$^{-1}$. 
The observation of optical polarization from  L dwarfs with 
known projected angular velocity implies non-sphericity of the photosphere
due to rotation.

 The oblateness of a rotating object has been discussed by \citet{cha33}
in the context of
polytropic gas configuration under hydrostatic equilibrium.  For a slow rotator,
the relationship for the oblateness $f$ of a stable
polytropic gas configuration under hydrostatic equilibrium is given by
\begin{eqnarray}\label{obl}
f=\frac{2}{3}C\frac{\Omega^2R^3_e}{GM},
\end{eqnarray}
where $M$ is the total mass, $R_e$ is the equatorial radius and $\Omega$ is
the angular velocity
of the object. $C$ is a constant whose value depends on the polytropic index.
For the polytropic index $n=0$, the density is uniform and  $C=1.875$.
This configuration is known as the Maclaurin spheroid.
For a polytropic index of $n=1.0$, $C=1.1399$, which is appropriate for
Jupiter \citep{hub84}. For non-relativistic completely degenerate gas,
$n=1.5$ and $C=0.9669$. 
The rotationally induced oblateness of solar planets has been discussed
in details by \citet{hub84} and \citet{mur00}.  Recently, the formalism
for oblateness
is extended to extra-solar planets by \citet{bar03} who used Darwin-Radau
relationship
 \begin{eqnarray}
f=\frac{\Omega^2R_e^3}{GM}\left[\frac{5}{2}\left(1-\frac{3}{2}K\right)^2+\frac{2}{5}\right]^{-1},
\label{darwin}
\end{eqnarray}
\citep{mur00, bar03}
to relate rotation to oblateness.
In eq.~\ref{darwin}, $K=I/MR_e^2 \leq 2/3$ is the moment of inertia parameter
of an object with moment of
inertia $I$. The Darwin-Radau relationship is exact for uniform density
objects ($K=0.4$)
and provides a reasonable (within a few percent of errors) estimation of the
oblateness of the solar planets.

Since L dwarfs have an extended convective region and their moment of
inertia is an undetermined parameter, we adopt the relationship given by
Chandrasekhar with the polytropic
index $n=1.0$ and $n=1.5$. $n<1.0$ would yield a too low density for brown
dwarfs, whereas  $n=1.5$ would provide minimum possible oblateness due to
rotation.
We have calculated the mass and radius of L dwarfs of different spectral types
by adopting the empirical relationship given in \citet{mar96}.

\section{The dust parameters}

The dust distribution in the atmosphere is calculated based on the one
dimensional cloud model of C03.  
This model assumes chemical equilibrium throughout the atmosphere, and uniform
density distribution across the surface of an object at each given pressure
and temperature. Under these assumptions, the number density of cloud particles
is given by

\begin{eqnarray}\label{density}
n(P)=q_c \left( \frac{\rho}{\rho_d}\right) 
\left(\frac{\mu_d}{\mu}\right)
\left(\frac{3}{4\pi a^3}\right),
\end{eqnarray}
where $\rho$ is the mass density of the surrounding gas, $a$ is the cloud
particle radius, $\rho_{d}$ is the mass density of the dust condensates,
$\mu$ and $\mu_d$ are the mean molecular
weight of atmospheric gas and condensates respectively.
The condensate mixing number ratio ($q_c$) is given as

\begin{eqnarray}
q_c=q_{below}\frac{P_{c,l}}{P}
\end{eqnarray}
for heterogeneously condensing clouds.
In the above equation, $q_{below}$ is the fraction of condensible vapor just
below the cloud base, $P_{c,l}$ is the pressure at the condensation
point (presented graphically in figure 1 of C03), and $P$ is the
gas pressure in the atmosphere. As in C03, we employ the ideal gas equation 
of state 
$\rho=\mu P/RT$
to relate
the gas mass density $\rho$ to the atmospheric temperature $T$ and pressure
$P$. $R$ is the universal gas constant and $\mu=2.36$ for solar composition
gas in which hydrogen is present in the molecular state.

The kinds of solid condensates possible in the  atmospheres of L dwarfs
include forsterite (Mg$_2$SiO$_4$), gehlenite (Ca$_2$Al$_2$SiO$_7$) and iron
(C03), as well as TiO$_2$ \citep{woi04}. However, forsterite forms in 
abundance in the atmosphere
and plays the most crucial role in governing the continuum spectrum. 
The other species such as gehlenite are much lees abundant than forsterite 
(about a factor of 10). In the present work, we have consider only forsterite.
The effect of other species on the degree of polarization is discussed later.
We have used the data from C03 for
forsterite : $\mu_d=140.7$ g mol$^{-1}$, $\rho_{d}=3.2$ g cm$^{-3}$,
and $q_{below}=3.2\times 10^{-5}$ (assuming solar abundance distribution
of the elements as discussed in C03). Throughout our
investigation, we have considered the wavelength $\lambda=0.850$ $\mu$m as
the observed polarimetric data presented by \citet{oso04}
is obtained  at I band centered at this wavelength. The object
2MASS J0036+1820 is observed by \citet{oso04} at R
Band (centered at 0.641 $\mu m$) as well as at I band (centered at 0.85
$\mu m$). The same object is observed by \citet{men02} at I band (centered at
0.768 $\mu m$). Therefore, for this object we model the degree of polarization
at different wavelengths. The real part of the refractive index is fixed at
1.65 and the imaginary part is taken by interpolating the data given in
\citet{sco96}.  It should be mentioned that 
the refractive index of amorphous condensates might differ under different
physical conditions. 

Apart from the calculation of the grain number density, the location of the
cloud in the atmosphere plays an important role in determining the amount of
polarization. The location of the cloud base for different atmospheric models
and different chemical species is determined by
the intersection of the T-P profile of the atmosphere model and the
condensation curve $P_{c,l}$ as prescribed in C03. Taking the condensation 
curve for forsterite, we determine the base of the
cloud for each spectral type, from L0 to L8. Figure~\ref{pbase} presents the 
atmospheric pressure height
at which the cloud base is situated for models with different spectral types.
As the effective temperature decreases from L0 to L8, the cloud base is pushed
deeper into the atmosphere.  According to the condensation curve given in C03,
the base of forsterite cloud for a L8 object ($T_{\rm eff} \simeq$1480 K) is
situated at about 10.0 bar pressure height when the
surface gravity of the object is assumed to be $10^5$ cm s$^{-2}$. 
For a similar model,
\citet{woi04} found the base of TiO$_2$ at 9.4 bar pressure height. 
Similarly, for a L4 ($T_{\rm eff}\simeq$1820 K) object, the forsterite cloud
base is found to be situated at 2.2 bar while \citet{woi04} found it to be at
2.7 bar for TiO$_2$ cloud.

\placefigure{figure2}

Theoretical investigation \citep{tsuji04a} claims that the thickness of the dust
clouds and hence the location of the cloud deck influences the spectral
energy distribution of L and T dwarfs. Also, it is suggested that the vertical
height of dust cloud may vary for a given $T_{eff}$ and surface gravity
\citep{knap04, tsuji04a}.
The degree of polarization too is strongly
dependent on the vertical height of the dust cloud and hence on the
location of the cloud deck.
In C03, the cloud deck is considered to be at one scale height above the base.
The condensation curve $P_{c,l}$
decreases exponentially with the decrease in the atmospheric temperature and 
the value of $P_{c,l}$
become negligibly small at $T=1600 K$.
For different atmospheric models, this
temperature is attained at different atmospheric pressure height. The position
of the cloud deck for different spectral types is also presented in figure 2.
From figure~\ref{pbase} we note that the thickness
of the cloud decreases as one goes from L8 to L0. In other words, the
thickness of the dust cloud decreases with the increase in effective
temperature for a fixed value of surface gravity. For
L dwarfs hotter than L2, the cloud is very thin, much less than one scale 
height. For a L8 object, the forsterite cloud deck is calculated
to be situated at about 4 bar pressure height.
For a similar atmospheric model, \citet{woi04} calculated the TiO$_2$ cloud
deck at 0.24 bar. For a L4 object the forsterite cloud deck in our model is 
at 0.5 bar pressure height while the TiO$_2$ cloud deck calculated by
\citet{woi04} for a similar object is at 0.1 bar.
Therefore the cloud thickness in our model is smaller than that of 
\citet{woi04}.  The number density of TiO$_2$ grains for L4 and L8 objects
is presented graphically in Woitke \& Helling (2004). 

At present, there is no convincing justification in favor of any specific 
form of the particle size distribution function. C03 considered a particle
size distribution function that is
consistent with the measurements of grain distribution attained in Earth's
water clouds while Ackerman \& Marley (2001) and Saumon et al. (2000)
considered a broad lognormal size distribution. In the present work,
we adopt the size distribution function given by the latter authors
which is expressed as :
\begin{eqnarray}
f(d)=\frac{d}{d_0}\exp\left[-\frac{\ln^2(d/d_0)}{\ln^2\sigma}\right]
\end{eqnarray}
where $d$ is the diameter of the particles and $d_0$ is the mean diameter,
$\sigma$ is fixed at
1.3 that provides good fit to the red spectrum of L dwarfs.  

\section{Results and discussion}

In a heterogeneously condensing cloud model with forsterite as the dominant
constituent, the number density of grains, the cloud base and its vertical 
scale height are fixed. The remaining free
parameters are the mean particle diameter $d_0$ and the oblateness induced
by rotation. As discussed by C03, the particle sizes significantly vary with
the effective temperature, surface gravity and vertical height of the
dust cloud. In the present
work we have calculated the degree of polarization for a fixed surface gravity
$g=10^5$ cm s$^{-1}$. Figure 4 shows how the degree of polarization is altered
with different mean particle diameter for objects with different spectral type
and hence for different effective temperature. In a multiple scattering
scenario, it is important to consider different mean particle size at different
pressure scale. However, in a single scattering, the photon scatters only
with one particle and therefore, for a fixed effective temperature, a fixed mean
particle size is sufficient.  This means that at every atmospheric 
pressure level, we use the lognormal distribution with particle mean diameter
$d_0$.

\placefigure{figure4}

We have included contributions to polarization by multipoles $l$=2, 4 and 6 in
eq.~\ref{pol}.  However, it is found that at optical wavelengths
and for the particle size needed to account the observed polarization,
contribution from $l=2$ is dominant over that of higher values for $l$. 

 The projected rotational velocity for several L dwarfs has been determined 
from observational data.  In the absence of any knowledge on the inclination
angle, the actual rotational velocity cannot be determined, yielding
uncertainties in the oblateness. In the present work we have considered 
rotational velocity of 15, 25, 30, 40 and 50 km s$^{-1}$. Figure~5 and figure~6 
show how the degree of polarization significantly increases with the increase
in rotational velocity.

A comparison of the degree of polarization presented in figure~5 and figure~6
shows that if the mean particle diameter is increased, 
the degree of polarization decreases significantly 
because with the increase in grain size, the grain number density reduces
according to eq.~\ref{density}.

\placefigure{figure3}

As mentioned in section 5, the location of the cloud base plays an important
role in determining the degree of polarization. As one moves from L0 to L8,
the cloud base goes deeper in the atmosphere
owing to the decrease in effective temperature. This makes the vertical size 
of the cloud larger (figure~2) and hence substantial increase in the total 
number of dust grains. In figure~3, we
show how the degree of polarization is altered when the cloud base is changed
but the particle number density remains unaltered.
For an object of spectral type L4.0 ($T_{\rm eff} \simeq$ 1820 K) with surface
gravity $g=10^{5}$ cm s$^{-2}$, the cloud base should be situated at an 
atmospheric temperature about 1750 K. Keeping the particle number density the 
same as that calculated for L4 object,
we calculated the degree of polarization by varying the location
of the cloud base. The degree of polarization is found to increase as the
location of the cloud base is pushed to the deeper
in the atmosphere where the temperature increases from 1600 K to 1800 K. 
This implies that as one moves from L8 to L0, that is from cooler to hotter
dwarfs, the degree of polarization
decreases owing to the decrease in the vertical height of the dust cloud.

\placefigure{figure4}

 As mentioned in section 1, Menard et al. (2002) for the first time, detected
linear polarization from three L dwarfs. They use I bessel filter having
the central wavelength at 0.768 $\mu m$. Out of the three objects that show
confirmed polarization, one object (DENIS-P J0255-4700) belongs to the
spectral type L8 and has high rotational velocity (40 km s$^{-1}$). This
object shows a linear polarization of 0.167\%. On the other hand, \citet{oso04} 
have reported confirmed polarization from nine L dwarfs in
I band (centered at 0.85 $\mu m$) and one L dwarf (2MASS J0036+1821)
in I as well as in R band (centered at
0.641 $ \mu m$). The object 2MASS J0036+1821 is also observed by Menard et al.
(2002) in I band but at different central wavelength. Hence, the degree of
polarization at three different wavelength region is available for this object. 

 Among the ten L dwarfs that are found to show confirmed polarization by
\citet{oso04}, one (2MASS J2244+20) is reported to be significantly redder than
those of other mid-L to late L dwarfs in near-infrared and infrared. Out of
the remaining nine L dwarfs, the projected rotational velocity of only four
objects is known. The object 2MASS J2252-17 belong to the spectral type L 7.5.
Although it's rotational velocity is not known, it shows degree of polarization
as high as 0.62 \%. Unless the rotational velocity of this object is very high
(even higher than the fastest L dwarf Kelu 1), the large difference in the
observed degree of polarization from this object and DENIS-P J0255-4700
cannot be explained as the mean particle size in the atmosphere of L dwarfs
of same spectral type should not differ much. Further theoretical investigation
on the distribution and location of condensates in the coolest L dwarfs are
needed before modeling their degree of polarization. Hence, in the present
work we discuss linear polarization of L dwarfs of spectral type ranging from
L0 to L6. 

\placefigure{figure5}
\placefigure{figure6}

Figure~5 shows the degree of linear polarization at $\lambda=0.850$ $\mu$m 
for objects with different spectral types but fixed surface gravity 10$^{5}$
cm s$^{-2}$. It is found that the observed degree of polarization of several
L dwarfs can be fitted if the mean diameter of grain is taken as
1.4 $\mu m$ and with the polytropic index n=1.0. The figure shows that the
change in polarization with the
spectral type is not linear but overall the degree of polarization decreases
as one moves from L1 to L6.  The degree of polarization decreases
substantially for objects hotter than L1 because 
condensation is not favored at such high temperature. The location of the 
cloud base shifts to deeper region of the atmosphere as the effective
temperature decreases and hence should cause an increase in the polarization
(figure~3).  However, the total amount of condensing material is conserved.
Therefore, as the scale height of the cloud layer becomes smaller, 
the particle number density becomes higher yielding into higher polarization.
Beyond L1 object, the temperature becomes too high to favor condensation and
the polarization falls rapidly to zero.

 \citet{oso04} detected the polarization in the Johnson R and
I-band filters centered on 0.641 and 0.850 $\mu m$ respectively with the
passband of these filters as 0.158 and 0.15 $\mu m$. Menard et al. (2002)
detected the polarization in the Bessel I-band filter on 0.768 $\mu m$ with
the passband as 0.138 $\mu m$. We have calculated the degree of polarization
at the central wavelengths. The change in degree of polarization over the
spread of wavelength should by and large be absorbed in the error bars.

Both 2MASS J1707+43 and 2MASS J1412+16 belong to the spectral type of L0.5.
2MASS J1412+16 having projected rotational velocity 16.4 kms$^{-1}$ shows
degree of polarization 0.57$\pm$0.19 while 2MASS J1707+43 shows degree of
polarization 0.23$\pm$0.06 but its rotational velocity is not known. Figure~5
shows that the observed data of these two objects
can well be explained if the rotational velocity of 2MASS J1707+43 is 
much less than 15 kms$^{-1}$. The model assumes the
mean diameter of grain is 1.4 $\mu m$ and the polytropic index n=1.0.
The same model can explain the observed polarization
of 2MASS J1507-16 (L5 with projected rotational velocity 27.2 kms$^{-1}$) if
its effective temperature is slightly higher then that given by the spectral
type - T$_{eff}$ polynomial formula. We put it as L5.5 to show that if the
effective temperature of this object corresponds to spectral type L5.5 in
the polynomial formula then the observed polarization can be well fitted.
The same model can explain the observed polarization from 2MASS J2158-15 (L4),
2MASS J0141+18 (L4.5) and 2MASS J0144-07 (L5) if their rotational velocities
are about 45, 25, 25 kms$^{-1}$ respectively. However, this model fails to
fit the data from Kelu 1 (L2.5) unless its actual rotational velocity is between
25 to 30 kms$^{-1}$.

Figure~6 presents the degree of polarization with the same model but with larger
mean particle diameter. It is found that the observed degree of polarization
from 2MASS 1707+43 (L0.5) can also be explained if its rotational velocity
is 20 kms$^{-1}$ but the atmosphere contains grains with mean diameter 3.0
$\mu m$. However, the mean grain size of objects with the same spectral type
should not differ much and therefore we predict the rotational velocity of
this object is about 5-10 kms$^{-1}$ with mean particle diameter 1.4 $\mu m$
as presented in figure~5. Figure~6 shows that the observed polarization of
Kelu 1 (L2.5) can be explained if the mean particle diameter is 3.0 $\mu m$.
However, it is worth mentioning here that the actual rotational velocity of L
dwarfs can not be determined from their projected rotational velocity unless
the inclination angle is known. Figure~6 shows that the model with mean particle
size 3.0 $\mu m$ and polytropic index n=1.5 ( non-relativistic completely
degenerate polytropic distribution) can explain the observed polarization of
2MASS J0141+18 (L4.5) and 2MASS J0144-07 (L5.0) if their rotational velocities
are the same of Kelu 1 (60 kms$^{-1}$).  

As mentioned earlier, the object 2MASS J0036+1821 (L3.5) is observed in three
different wavelength regions and the degree of polarization is found to
decrease substantially with the increase in wavelength. This trends strongly
supports the presence of dust in the atmosphere of L dwarfs as it is very much
unlikely that any other mechanisms such as the presence of magnetic field can
explain this. Figure~7 presents the degree of polarization as a function of
wavelength for L3.5 object with the polytropic index n=1.5 and the rotational
velocity v=15 kms$^{-1}$. It is obvious from the observed data at three
different wavelengths that the atmosphere of 2MASS J0036+1821 should have
grains of sub-micron size. We find the best fit of the three observed data
with the mean particle diameter $d_0=$0.43 $\mu m$. Larger grain size would
have made the polarization to peak at longer wavelength. 

 The degree of polarization increases with the decrease in the polytropic index 
because with $n=1.0$, the oblateness of the object is higher than that with
$n=1.5$ for the same rotational velocity and surface gravity. On the other
hand Sengupta (2003) showed that the oblateness decreases with the increase
in surface gravity yielding into less amount of polarization.

As mentioned before, we have considered only forsterite in our models as it is
very common in the atmosphere of L type dwarfs with solar metalicity. This is
because of large abundances of Mg, Si and O. However, gehlenite, enstatite
etc. should also be present in the atmosphere in fairly good abundance. The
condensation curve presented in C03 shows that the base of gehlenite is
situated much deeper in the atmosphere than that of forsterite yielding into
a larger verticle size of the cloud. Further, iron with higher refractivity
may undercircle most of the silicate clouds. Inclusion of all these will
lead to substantial increase in degree of polarization and therefore, much
smaller grain size may be needed in order to explain the observed degree of
polarization. 
  
Lastly, multiple scattering will lead to much less polarization
and therefore in order to fit the observed polarization one has to either
consider much higher oblateness of the objects or has to increase the grain
number density substantially. An increase in grain number density needs
smaller grain size which may contradict the present theoretical understanding
on the nature and formation of dust grain. On the other hand, differential
photometric observation of several L dwarfs could not detect any non-periodic
variability of many objects that show high polarization. This indicates 
optically thin dust shell and therefore polarization by single dust scattering
is quite reasonable. 

\section{Conclusions}

We have investigated the optical linear polarization from L dwarfs of 
different spectral type by considering single dust scattering. Forsterite is
considered to be the dominant species among
the various condensates that could be present in the atmosphere of L dwarfs.
The location of the cloud base and the cloud deck is determined from the
condensation curve for forsterite and the atmospheric temperature-pressure 
profiles of different spectral types. The surface gravity
is fixed at $10^5$ cm s$^{-2}$ and a wide range of rotational velocity is
considered. It is found that the degree of linear polarization decreases from
hotter to cooler L dwarfs. However, L dwarfs
with effective temperature greater than 2200 K should not show detectable
amount of polarization due to dust scattering as most of the dust would either
evaporate from the atmosphere or condensation is not favored at such high
temperature.  It is found that the mean 
diameter of grains that is consistent with the observed polarization should
not exceed a few micron although a small amount of very large grains
at the base of the cloud for comparatively cooler L dwarfs may well be 
accommodated. However, the observational data of 2MASS J0036+1821 clearly
indicates the presence of sub-micron size grain.
Further polarimetric observation at the optical and at other wavelengths
would provide convincing information
on the properties and distribution of dust in the atmosphere of L dwarfs.

\acknowledgments

We are thankful to the referee for several valuable comments and suggestions.

\clearpage

\begin{figure}
\plotone{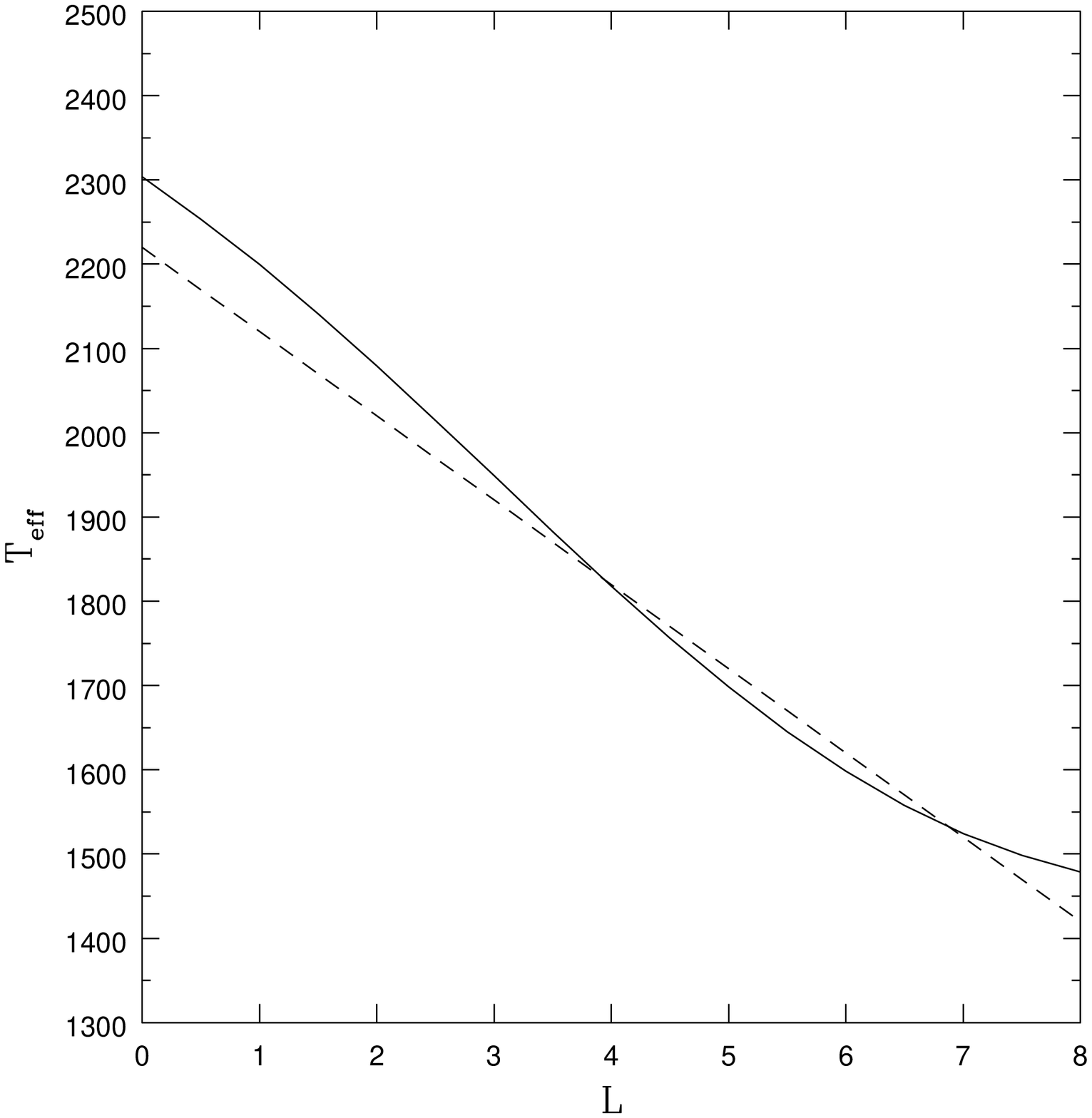}
\caption{Effective temperature vs spectral type : solid line represents the
calibration of $T_{eff}$ using the sixth order polynomial fit as given in
\citet{goli04} while the dashed line represents that by using the 
linear relationship given in Stephens et al. (2001).}
\end{figure}

\begin{figure}
\plotone{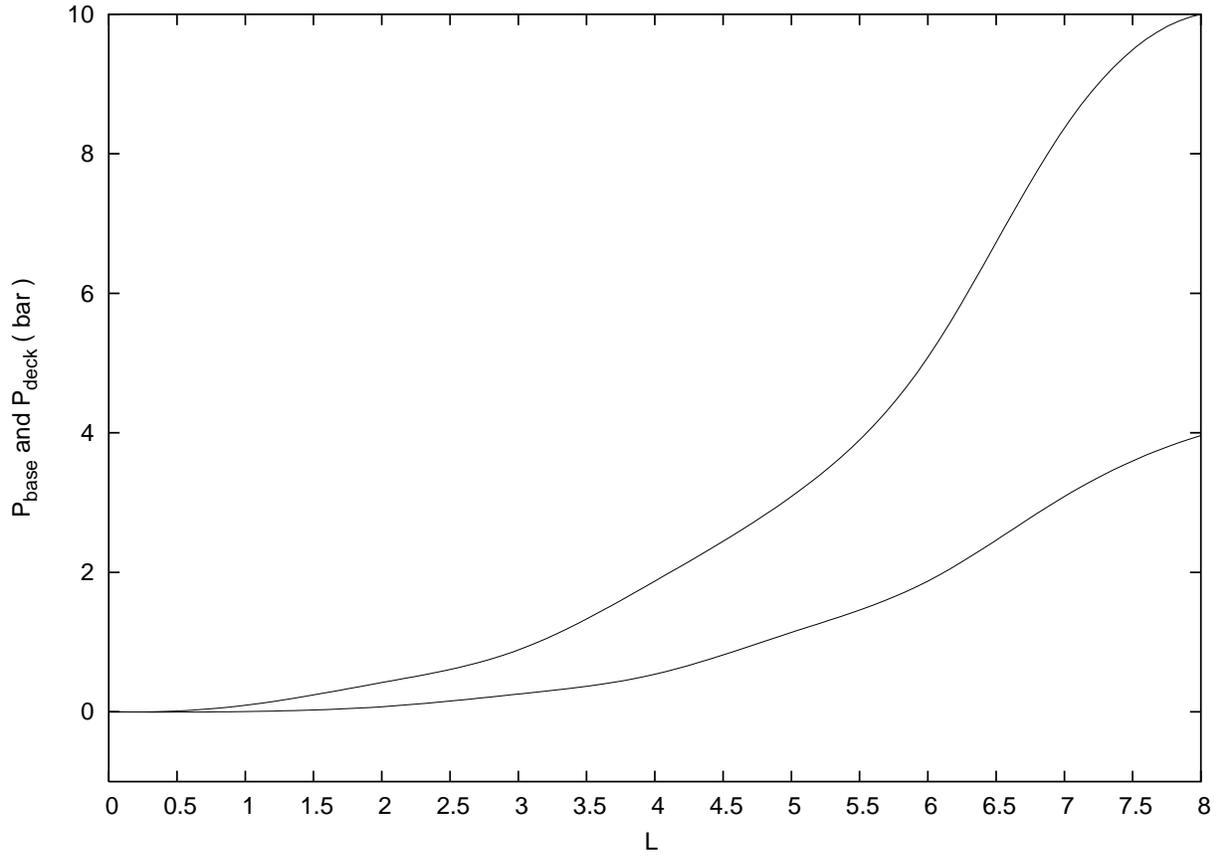}
\caption{Locations of forsterite cloud base (upper curve) and cloud deck (lower
curve) for different spectral type L.}
\label{pbase}
\end{figure}

\begin{figure}
\plotone{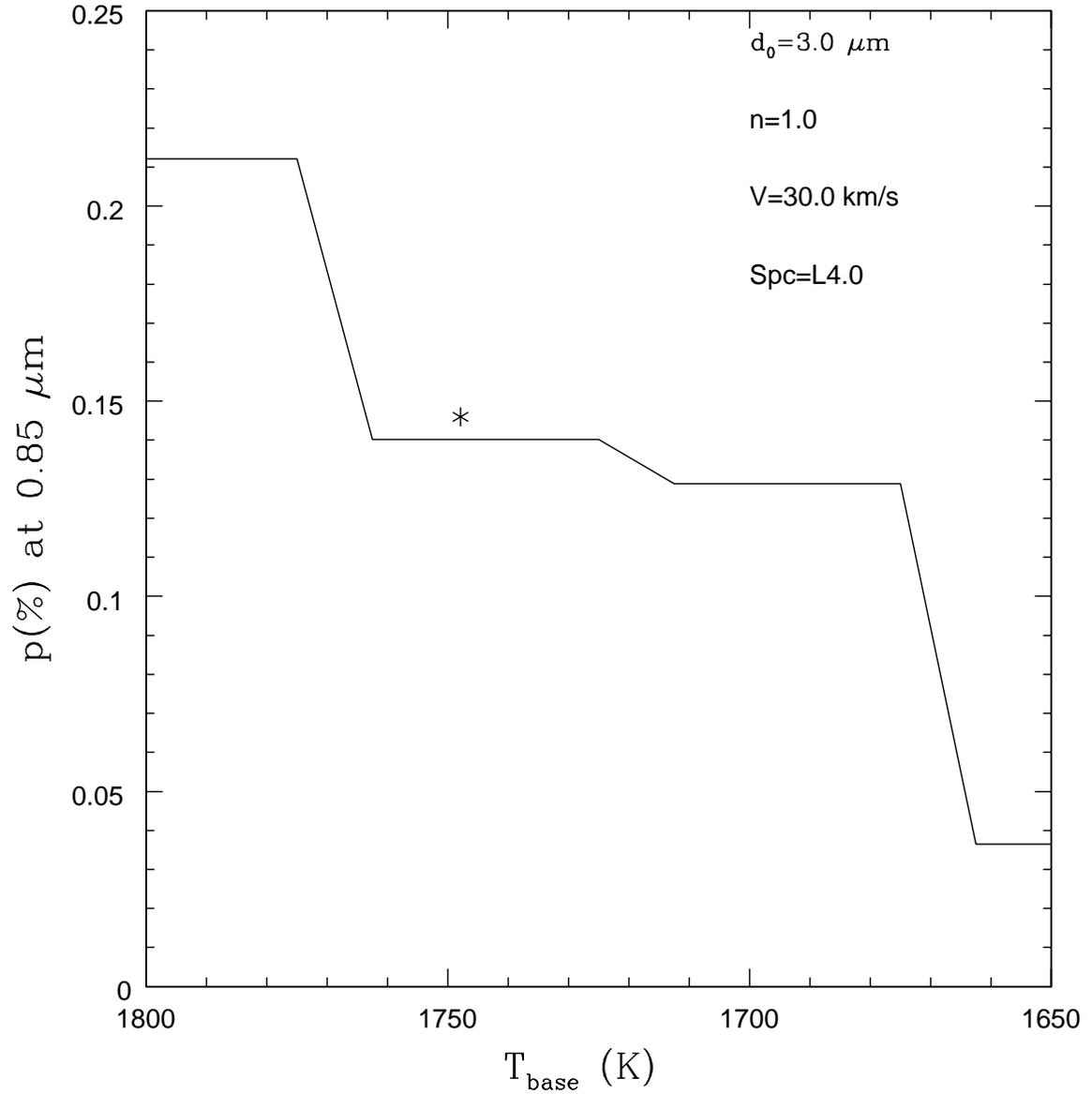}
\caption{Degree of polarization for L4.0 object with the cloud base situated
at different temperature in the atmosphere but with a fixed particle number
density. The star indicates the degree of
polarization when the cloud base is situated at 1750 K as determined from the
intersection of the condensation curve and the T-P curve for L4.0 with
$g=10^5$ cm s$^{-2}$.} 
\label{cbase}
\end{figure}

\begin{figure}
\plotone{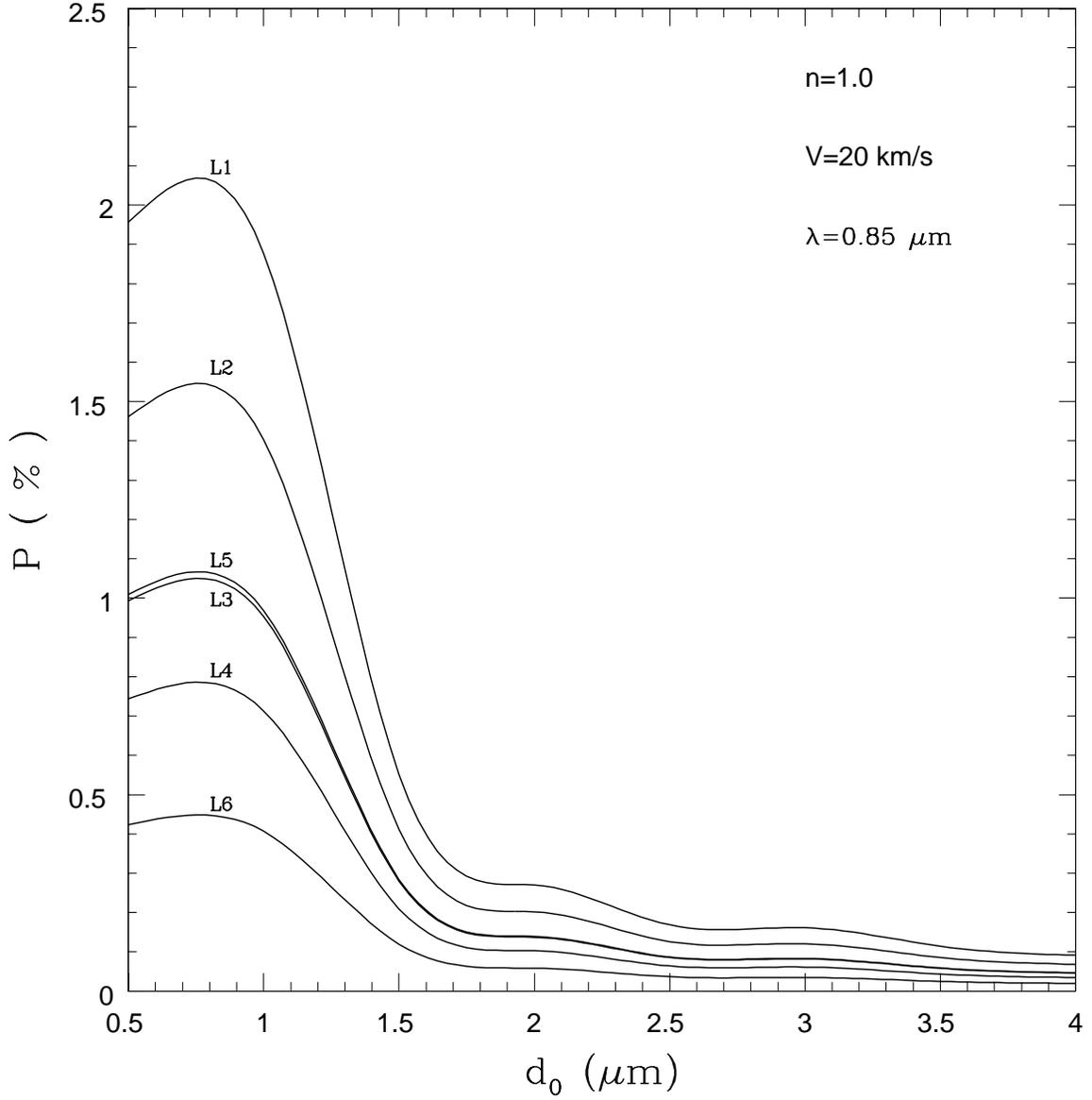}
\caption{Degree of polarization as a function of mean grain diameter $d_0$
for L dwarfs with different spectral type (L1-L6).}
\end{figure}

\begin{figure}
\plotone{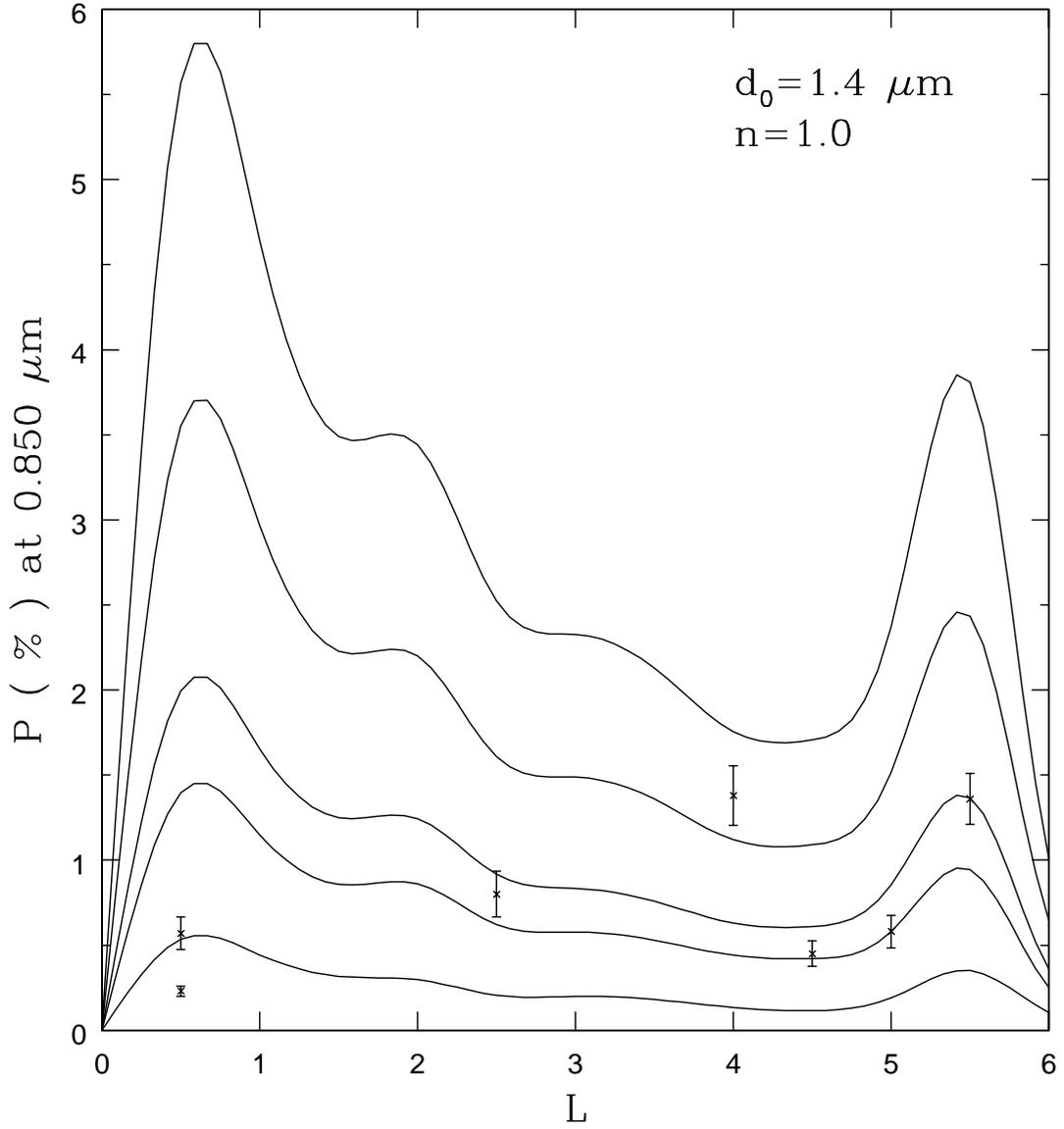}
\caption{Degree of polarization for objects with different spectral type with
the mean grain diameter $d_0=1.4 \mu$m and the polytropic index n=1.0.
The five curves are for rotational
velocities of  V=15, 25, 30, 40 and 50 km s$^{-1}$, from bottom to top. 
The observed polarization of  seven L dwarfs are plotted with their 
respective error bars.}\label{spd4}
\end{figure}

\begin{figure}
\plotone{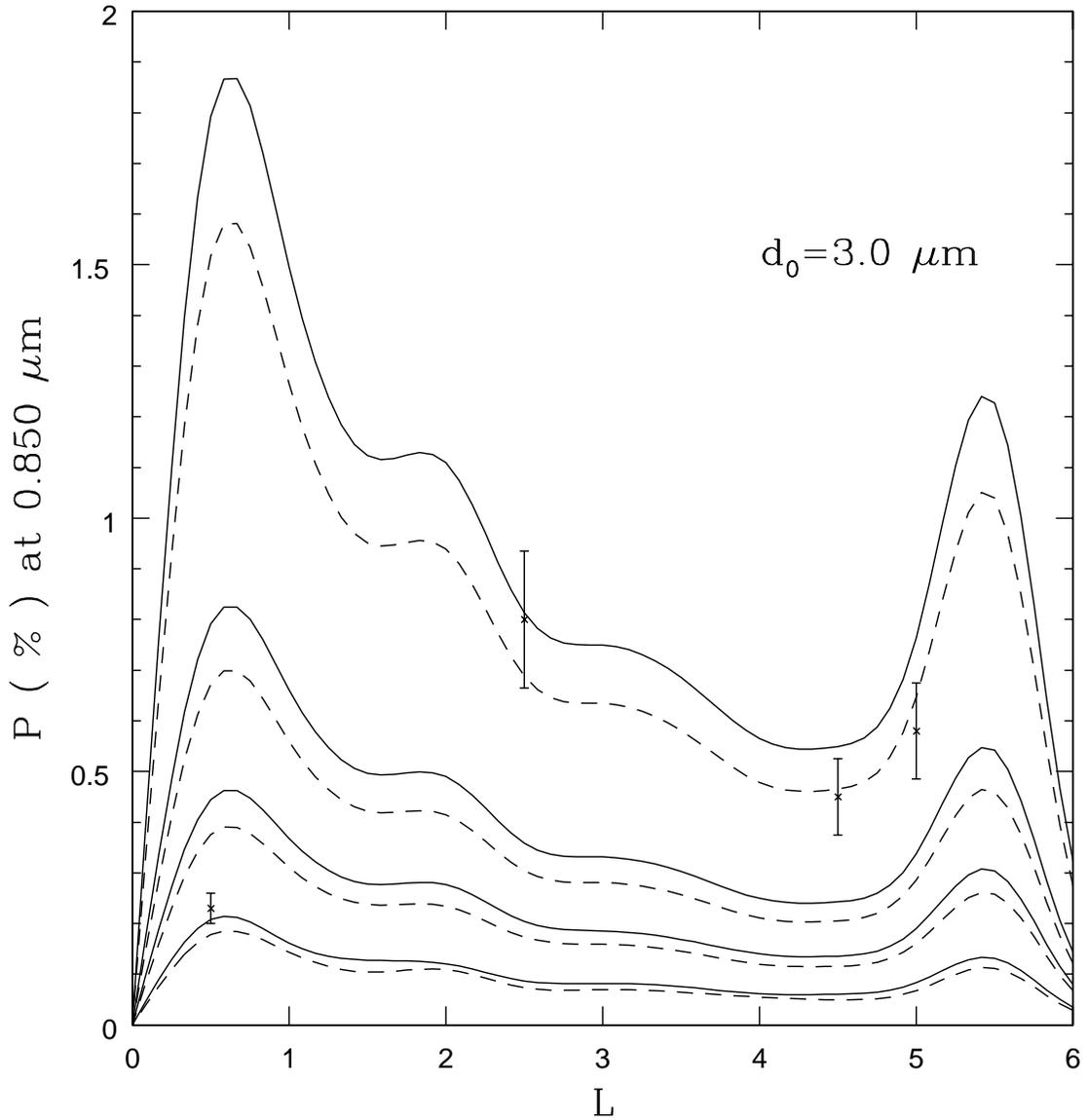}
\caption{Same as figure 4 but with the mean grain diameter $d_0=3.0 \mu m$.
The solid curves represent the degree of polarization with the polytropic index
n=1.0 while the dashed curves represent that with n=1.5. The four pair of
curves are for rotational velocities of V=20, 30, 40 and 60 km s$^{-1}$ from
bottom to top. The observed polarization of four L dwarfs are also presented.} 
\label{spd4a}
\end{figure}

\begin{figure}
\plotone{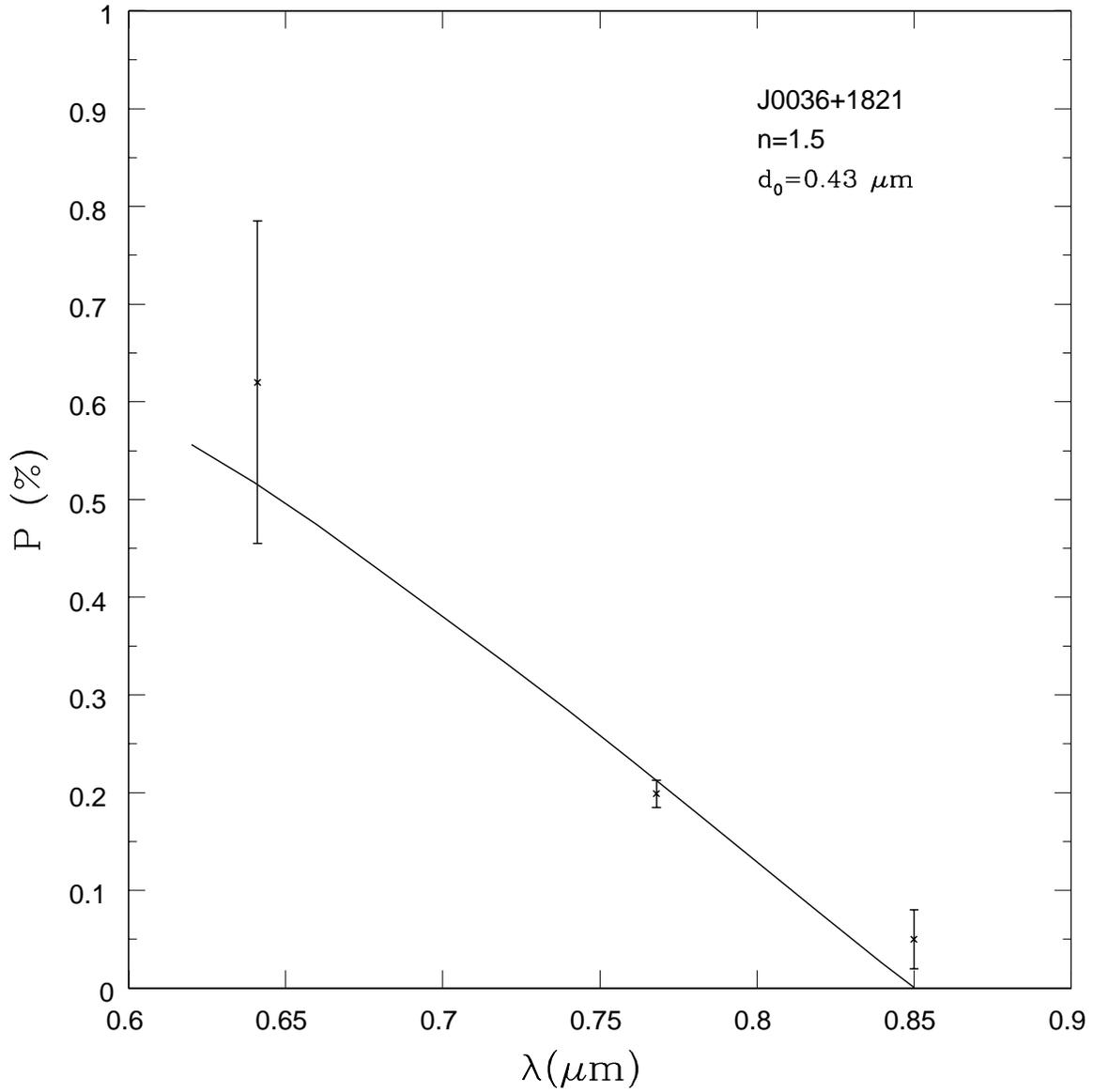}
\caption{Degree of polarization as a function of wavelength with the mean
grain diameter $d_0=0.43 \mu m$ and polytropic index n=1.5. The observed 
polarization of the L dwarf 2MASS J0036+1821 at three different wavelengths
are fitted with this model. The rotational velocity of this object is taken
as V=15 km s$^{-1}$}
\label{spd6}
\end{figure}

\end{document}